\documentclass[preprint,showpacs,preprintnumbers,amsmath,amssymb]{revtex4}
\usepackage{graphicx}
\usepackage{epsfig}
\usepackage{amsmath}
\usepackage{ascmac} 
\usepackage{graphics}
\usepackage{ascmac}
\usepackage{color}
\usepackage{amsmath,amssymb} 
\begin{document} 
\title{Green's function formalism for a condensed Bose gas consistent with infrared-divergent longitudinal susceptibility and Nepomnyashchii--Nepomnyashchii identity}
\author{Shohei Watabe$^{1}$} 
\author{Yoji Ohashi$^{2}$} 
\affiliation{$^1$ Department of Physics, The University of Tokyo, Tokyo 113-0033, Japan,
\\
$^2$ Faculty of Science and Technology, Keio University, Yokohama 223-8522, Japan.}
\begin{abstract} 
We present a Green's function formalism for an interacting Bose--Einstein condensate (BEC) satisfying the two required conditions: (i) the infrared-divergent longitudinal susceptibility with respect to the BEC order parameter, and (ii) the Nepomnyashchii--Nepomnyashchii identity stating the vanishing off-diagonal self-energy in the low-energy and low-momentum limit. These conditions cannot be described by the ordinary mean-field Bogoliubov theory, the many-body $T$-matrix theory, as well as the random-phase approximation with the vertex correction. In this paper, we show that these required conditions can be satisfied, when we divide many-body corrections into singular and non-singular parts, and separately treat them as different self-energy corrections. The resulting Green's function may be viewed as an extension of the Popov's hydrodynamic theory to the region at finite temperatures. Our results would be useful in constructing a consistent theory of BECs satisfying various required conditions, beyond the mean-field level. 
\end{abstract} 
\maketitle
\section{Introduction}\label{SecI} 
\par 
The Bogoliubov-type mean-field theory~\cite{Bogoliubov1947} has successfully clarified various superfluid phenomena of ultracold Bose gases. However, the theory of Bose--Einstein condensates (BECs) still has room for improvement. The Hartree-Fock-Bogoliubov (HFB) approximation gives a finite energy gap~\cite{Griffin1996}, and the HFB-Popov (Shohno) theory~\cite{Popov1964A,Popov1964B,Popov1983BOOK,Griffin1996,Shi1998,Shohno1964} unphysically concludes the first-order phase transition~\cite{Shi1998,Reatto1969} (whereas a real Bose gas is expected to exhibit the second-order phase transition). These mean-field type BEC theories also assume a static self-energy $\Sigma$, where its off-diagonal self-energy part $\Sigma_{12}$ characterized by two outgoing particle lines is specific to the BEC phase. This static result contradicts with the exact identity proved by Nepomnyashchii and Nepomnyashchii~\cite{Nepomnyashchii1975,Nepomnyashchii1978}, stating the vanishing off-diagonal self-energy $\Sigma_{12}$ in the low-energy and low-momentum limit. 
\par 
When we try to go beyond the mean-field approximation to include many-body correlations, we suffer from the infrared divergence associated with fluctuations of BECs. The infrared singularity directly appears in some quantities such as the correlation functions of the phase and amplitude fluctuations of a BEC order parameter~\cite{Patashinskii1973,Nepo1983,Weichman1988,Giorgini1992,Griffin1993BOOK,Castellani1997,Zwerger2004,Sinner2010,Dupuis2011,Podolsky2011} (that are also referred to as the transverse and longitudinal response functions in the literature, respectively). On the other hand, in some cases such as the density-density correlation function~\cite{Gavoret1964}, the infrared divergence does not appear in the final result. The singularity only appears on the way of calculation. Indeed, the density response function satisfies the compressibility sum-rule~\cite{Gavoret1964}. One thus needs to carefully treat the infrared divergence, depending on what we are considering. 
\par 
For curing the infrared divergences in BEC theories, a number of ideas have been proposed. Instead of bosonic fields, hydrodynamic variables (such as density and phase) have been adopted to describe BECs in the low momentum region at $T=0$~\cite{Popov1972,Popov1979,Popov1983BOOK}. This so-called Popov's hydrodynamic approach correctly describes the long-range correlations. A renormalization group technique has also been applied to the BEC phase at $T=0$~\cite{Castellani1997,Pistolesi2004,Dupuis2007,Sinner2010}. To obtain correct infrared behaviors in this approach, the Ward--Takahashi identities associated with the gauge symmetry play an important role. The infrared divergences are also removed by an artificial field that breaks $U(1)$ gauge symmetry~\cite{Capogrosso-Sansone2010}. In this approach, one takes the limit of vanishing symmetry breaking terms after calculating physical quantities, and long-range correlations are corrected by incorporating the Popov's hydrodynamic theory. 
\par 
In this paper, we construct a Green's function formalism that can correctly describe the low-energy singularity of the longitudinal response function $\chi_\parallel({\bf p},\omega)$. This function is a typical quantity exhibiting the infrared divergence below the BEC phase transition temperature $T_{\rm c}$. This infrared behavior is strongly related to the so-called Nepomnyashchii--Nepomnyashchii (NN) identity~\cite{Nepomnyashchii1975,Nepomnyashchii1978}. Indeed, it has been shown that $\chi_\parallel(0,0)$ is proportional to $\Sigma_{12}^{-1}(0,0)$~\cite{Nepo1983}. 
\par 
The longitudinal response function is also a useful quantity in constructing a consistent theory of BECs. The Bogoliubov mean-field theory incorrectly gives a finite value of $\chi_\parallel(0,0)$~\cite{Nepo1983,Weichman1988}. This approximation only includes fluctuations around the mean-field order parameter to the second-order, where longitudinal (amplitude) and transverse (phase) fluctuations of the BEC order parameter are decoupled from each other in the Bogoliubov Hamiltonian. Anharmonic effects beyond such a Gaussian approximation have been pointed out to be important for the NN identity~\cite{Weichman1988}. However, the so-called many-body $T$-matrix theory, which involves interaction effects beyond the mean-field level, still cannot reproduce the infrared singularity of the longitudinal response function, nor the NN identity~\cite{Shi1998,Watabe2013}. 
\par 
We note that the infrared divergence of the longitudinal susceptibility associated with an order parameter is a general phenomenon in a system with spontaneously broken continuous symmetry. This divergence was originally discussed in a Heisenberg ferromagnet~\cite{Vaks1968}, and was extended to a general system described by a multi-component ordering field~\cite{Patashinskii1973}. Although the longitudinal susceptibility has not been observed in an ultracold Bose gas, the singularity of the longitudinal dynamical susceptibility is observable in Bose--Einstein condensation of magnons in a quantum Heisenberg antiferromagnet via neutron scattering~\cite{Kreisel2007}. 
\par 
Here, we explain our strategy in this paper. Effects of a particle-particle interaction can be conveniently included in the single-particle thermal Green's function $G$ through the Dyson equation 
\begin{equation} 
G({\bf p},i\omega_n) = {1 \over G^{-1}_0({\bf p},i\omega_n)-\Sigma({\bf p},i\omega_n)}. 
\label{eq.intro1}
\end{equation} 
Here, $G_0 $ is the Green's function for a free Bose gas, $\Sigma $ is the irreducible self-energy, and $\omega_n$ is the boson Matsubara frequency. Equation (\ref{eq.intro1}) is the most conventional expression in considering an interacting Bose gas. This equation is actually a $(2\times 2)$-matrix in the BEC phase. 
\par 
One may also include many-body corrections into $G$ through the reducible self-energy $\Sigma '$ by using the expression 
\begin{equation}
G({\bf p},i\omega_n)=G_0({\bf p},i\omega_n) + G_0({\bf p},i\omega_n) \Sigma ' ({\bf p},i\omega_n)G_0({\bf p},i\omega_n). 
\label{eq.intro2}
\end{equation} 
The reducible self-energy $\Sigma '$ is related to the irreducible self-energy $\Sigma$ through 
\begin{equation}
\Sigma ' ({\bf p},i\omega_n)={1 \over 1-\Sigma({\bf p},i\omega_n)G_0({\bf p},i\omega_n)} \Sigma({\bf p},i\omega_n).
\label{eq.intro3} 
\end{equation} 
Equations~(\ref{eq.intro1}) and (\ref{eq.intro2}) are equivalent to each other. 
\par 
We may also employ the hybrid version of Eqs.~(\ref{eq.intro1}) and (\ref{eq.intro2}), given by 
\begin{equation}
G({\bf p},i\omega_n)={\tilde G}({\bf p},i\omega_n) + {\tilde G}({\bf p},i\omega_n){\tilde \Sigma}({\bf p},i\omega_n){\tilde G}({\bf p},i\omega_n), 
\label{eq.intro4}
\end{equation} 
where we divide the self-energy $\Sigma$ into two parts, i.e., $\Sigma =\Sigma_{\rm a} +\Sigma_{\rm b}$. In Eq.~(\ref{eq.intro4}), $\Sigma_{\rm a}$ is treated as the self-energy correction to $G_{0}$, which provides ${\tilde G}^{-1}=G_0^{-1}-\Sigma_{\rm a}$. On the other hand, $\tilde \Sigma$ is given by Eq.~(\ref{eq.intro3}), where $\Sigma$ and $G_0$ are replaced by $\Sigma_{\rm b}$ and ${\tilde G}$, respectively. If the sum of $\Sigma_{\rm a}$ and $\Sigma_{\rm b}$ provides the exact irreducible self-energy possessing all orders of interactions, Eq.~(\ref{eq.intro4}) is a rewriting of Eq.~(\ref{eq.intro1}). 
\par 
In most cases, we need an approximate treatment of many-body effects. In an extreme case, one may introduce different approximations between ${\tilde G}$ in the first term of (\ref{eq.intro4}) and those in the second term, giving the form 
\begin{equation}
G({\bf p},i\omega_n)={\tilde G}({\bf p},i\omega_n) + \tilde G_{\rm L}({\bf p},i\omega_n){\tilde \Sigma}({\bf p},i\omega_n) \tilde G_{\rm R}({\bf p},i\omega_n).
\label{eq.intro5}
\end{equation} 
Equation~(\ref{eq.intro5}) is more flexible than Eq.~(\ref{eq.intro1}) in the sense that one may employ different approximations in the first and the second terms. This flexibility is particularly useful in constructing the BEC theory that satisfies various required conditions, such as the infrared divergence of the longitudinal response function, the NN identity, the Hugenholtz-Pines relation~\cite{Hugenholtz1959} as well as the second-order phase transition. 
\par
In this paper, we employ the hybrid expression in Eq.~(\ref{eq.intro5}). We treat the first term in Eq.~(\ref{eq.intro5}) within the many-body $T$-matrix approximation (MBTA), as well as the random phase approximation (RPA) with the vertex correction. In a previous paper~\cite{Watabe2013}, we examined a weakly interacting Bose gas within the framework of the ordinary Green's function formalism in Eq.~(\ref{eq.intro1}). Evaluating the self-energy within the MBTA as well as the RPA with the vertex correction, we found that these many-body theories describe the enhancement of $T_{\rm c}$ as predicted by various methods~\cite{Andersen2004}, whereas they do not meet the NN identity. We overcome this problem in this paper, by determining the second term in Eq.~(\ref{eq.intro5}) so as to cure the broken NN identity. The resulting Green's function is found to also reproduce the infrared divergence of the longitudinal response function, as well as the Hugenholtz-Pines relation. 
\par 
We determine ${\tilde \Sigma}$ in Eq.~(\ref{eq.intro5}) on the basis of the hydrodynamic theory developed by Popov~\cite{Popov1972,Popov1979,Popov1983BOOK}. Indeed, our approach based on Eq.~(\ref{eq.intro5}) is strongly related to the Popov's hydrodynamic theory. The Green's function in Eq.~(\ref{eq.intro5}) has formally the same structure as that given in the Popov's hydrodynamic theory. In this hydrodynamic theory, a factor corresponding ${\tilde \Sigma}$ in Eq.~(\ref{eq.intro5}) exhibits infrared divergence that originates from phase fluctuations of the BEC order parameter. Using this, Popov obtained the vanishing off-diagonal self-energy in the low-energy and low-momentum limit (NN identity). This result provides a crucial key in determining ${\tilde \Sigma}$. 
\par 
Section~\ref{SecII} presents our Green's function formalism. In Sec.~\ref{SecIII}, we examine low-energy properties of the longitudinal response function in our formalism. We explicitly show that our formalism satisfies the NN identity. We also discuss how our approach is related to the Popov's hydrodynamic theory. In Sec.~\ref{SecIV}, we examine the condensate fraction as a function of the temperature, to see how the present theory affects the previous results based on the ordinary Green's function formalism in Eq.~(\ref{eq.intro1}). Throughout this paper, we set $\hbar = k_{\rm B} = 1$, and the system volume $V$ is taken to be unity. 
\par 
\section{Framework}\label{SecII}
\par 
We consider a three-dimensional Bose gas with an atomic mass $m$. The Hamiltonian is given by 
\begin{equation}
H=\sum_{\bf p} (\varepsilon_{\bf p} - \mu) a_{\bf p}^{\dag} a_{\bf p} + \frac{U}{2} \sum_{{\bf p},{\bf p}',{\bf q}} a_{{\bf p} + {\bf q}}^{\dag} a_{{\bf p}'-{\bf q}}^{\dag} a_{{\bf p}'} a_{{\bf p}}, 
\label{eq1}
\end{equation} 
where $a_{\bf p}$ is the annihilator of a Bose atom with the kinetic energy $\varepsilon_{\bf p}-\mu = {\bf p}^{2}/(2m)-\mu$, measured from the chemical potential $\mu$. We consider a weak repulsive interaction $U (> 0)$, which is related to the $s$-wave scattering length $a$ as
\begin{align} 
\frac{4 \pi a}{m} = \frac{U}{1+ \displaystyle{U} \sum_{{\bf p}}^{p_{\rm c}}  \displaystyle{ \frac{1}{2 \varepsilon_{\bf p}} }  }, 
\label{eq3}
\end{align} 
where $p_{\rm c}$ is a cutoff momentum.
\par 
The BEC phase is conveniently characterized by the BEC order parameter $\langle a_{{\bf p}=0}\rangle$. It is related to the condensate fraction $n_0$ through $\langle a_{{\bf p}=0}\rangle=\sqrt{n_0}$~\cite{Bogoliubov1947}. In this paper, we take $\langle a_{{\bf p}=0}\rangle$ as a real number, without loss of generality.
\par 
We consider the $(2\times 2)$-matrix single-particle thermal Green's function having the form in Eq.~(\ref{eq.intro5}). We divide a self-energy $\Sigma({\bf p},i\omega_n)$ into the sum of the singular part $\Sigma^{\rm IR}({\bf p},i\omega_n)$ (which exhibits infrared divergence) and the regular part $\Sigma^{\rm R}({\bf p},i\omega_n)$ (which remains finite even in the low-energy and low-momentum limit). For ${\tilde G}$ and ${\tilde \Sigma}$ in Eq.~(\ref{eq.intro5}), we take
\begin{equation}
{\tilde G}({\bf p},i\omega_n)= {1 \over i\omega_n\sigma_3- \varepsilon_{\bf p}+\mu-{\Sigma}^{\rm R}({\bf p},i\omega_n)}, 
\label{eq.self1}
\end{equation} 
\begin{equation}
{\tilde \Sigma}({\bf p},i\omega_n) = \Sigma^{\rm IR}({\bf p},i\omega_n).
\label{eq.self2}
\end{equation} 
Here, $\sigma_i$ ($i=1,2,3$) are Pauli matrices. We determine the chemical potential $\mu$, so as to satisfy the Hugenholtz-Pines relation $\mu = \Sigma^{\rm R}_{11}(0,0) - \Sigma^{\rm R}_{12}(0,0)$~\cite{Hugenholtz1959}. 
\par
\begin{figure}
\begin{center}
\includegraphics[width=9cm]{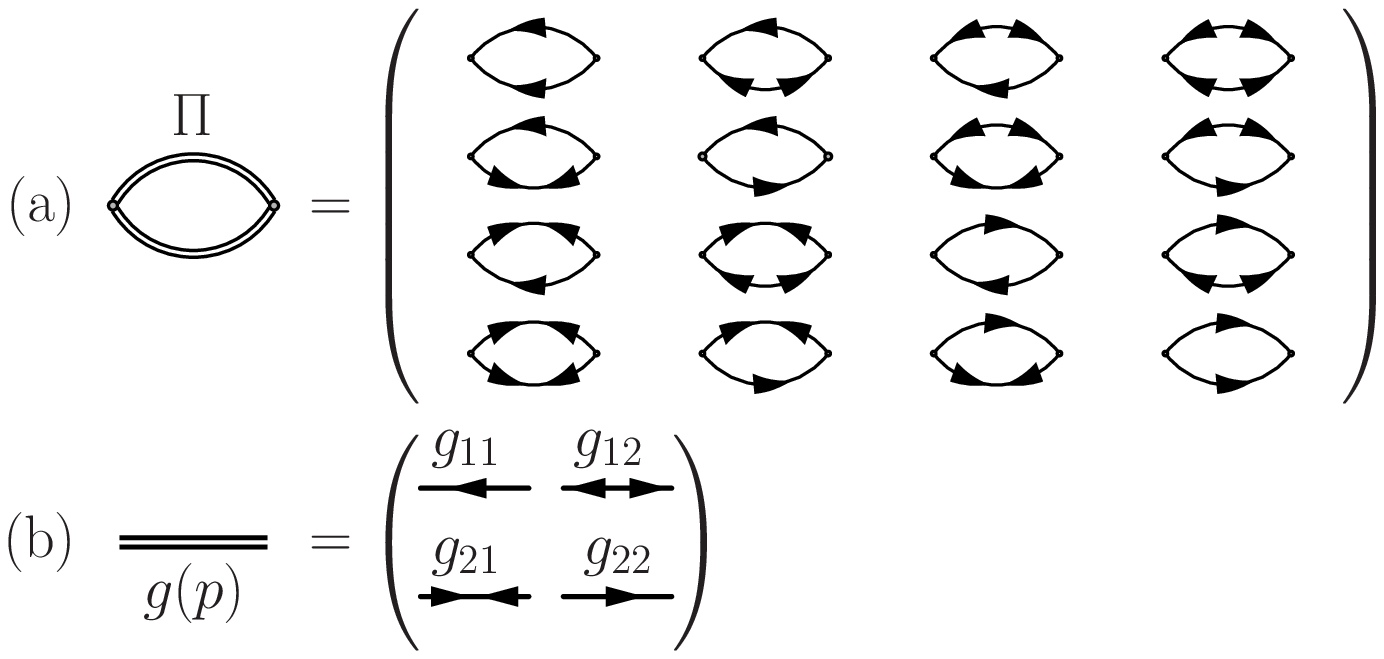}
\end{center}
\caption{(a) Generalized polarization function $\Pi$. (b) Single particle Green's function $g(p)$ used in (a). }
\label{fig1.fig}
\end{figure}

  
To explain how to divide the self-energy into two parts in the the MBTA and the RPA, we conveniently introduce the $(4\times 4)$-matrix generalized correlation function~\cite{Watabe2013} 
\begin{eqnarray}
\Pi (p) = -T\sum_q g(p+q) \otimes g(-q), 
\label{2bdyGreen}
\end{eqnarray} 
where $\otimes$ is the Kronecker product. Here, we have used the simplified notation $p=({\bf p},i\omega_n)$. The correlation function $\Pi(p)$ is diagrammatically given in Fig.~\ref{fig1.fig} (a). In Eq.~(\ref{2bdyGreen}), $g_{} (p)$ is the $(2\times 2)$-matrix single-particle Green's function in the HFB--Popov approximation, given by 
\begin{align}
g(p) = \frac{ 1 }{i \omega_{n} \sigma_{3} - \xi_{\bf p}  - U n_{0} \sigma_{1}}, 
\label{EqE1}
\end{align} 
where $\xi_{\bf p} = \varepsilon_{\bf p} + U n_{0}$. Using the symmetry properties $g_{22}(p) = g_{11} (-p)$ and $g_{12}(p) = g_{12}(-p)$, we reduce Eq.~(\ref{2bdyGreen}) to 
\begin{align}
\Pi(p) = & 
\begin{pmatrix}
\Pi_{11} (p) & \Pi_{12} (p) & \Pi_{12} (p) & \Pi_{14} (p) \\
\Pi_{12} (p) & \Pi_{22} (p) & \Pi_{14} (p) & \Pi_{12}^{*} (p) \\
\Pi_{12} (p) & \Pi_{14} (p) & \Pi_{22} (p) & \Pi_{12}^{*} (p) \\
\Pi_{14} (p) & \Pi_{12}^{*} (p) & \Pi_{12}^{*} (p)& \Pi_{11}^{*}  (p)
\end{pmatrix}. 
\label{RPAPiMatrix}
\end{align} 
The detailed expressions of $\Pi_{ij}$ are summarized in Appendix~\ref{AppendixA1}. 
\par 
We divide (\ref{RPAPiMatrix}) into the sum $\Pi(p)=\Pi^{\rm IR}(p)+\Pi^{\rm R}(p)$ of the singular part $\Pi^{\rm IR}(p)$ (which exhibits infrared divergence) and the regular part $\Pi^{\rm R}(p)$ (which remains finite in the low-energy and low-momentum limit). In this case, the singular part $\Pi^{\rm IR}(p)$ can be written so as to be proportional to $\Pi_{14}(p)$, giving the form 
\begin{align}
{ \Pi}^{\rm IR} (p) = \Pi_{14} (p) \hat C, 
\label{eq.secIII13}
\end{align}
with  
\begin{align}
\hat C = 
\begin{pmatrix}
1 & - 1 & -1 & 1 \\
- 1 & 1 & 1 & - 1 \\
- 1 & 1 & 1 & - 1 \\
1 & - 1 & -1 & 1
\end{pmatrix}. 
\end{align} 
The infrared singularity of (analytic continued) $\Pi_{14}$ is given by~\cite{Gavoret1964,Nepo1983,Griffin1993BOOK,Dupuis2011,Podolsky2011,Giorgini1992,Kreisel2007,Dupuis2011LP}
\begin{align}
\Pi_{14}({\bf p},i\omega_n\to\omega+i\delta) \propto 
\left \{ 
\begin{array}{ll}
\ln ( c_{0}^{2} {\bf p}^{2} - \omega^{2}) & (T = 0)
\\
1/|{\bf p}| &  (T \neq 0) . 
\end{array}
\right .
\label{eq10}
\end{align} 
(For the derivation of (\ref{eq10}), see Appendix~\ref{AppendixA2}.) In Eq.~(\ref{eq10}), $c_{0} = \sqrt{n_{0} U/m}$ is the Bogoliubov sound speed, and $\delta$ is an infinitesimally small positive number. 
\par 
The singular part $\Pi^{\rm IR}$ only appears in the BEC phase below $T_{\rm c}$. Indeed, the singular part $\Pi_{14}$ is constructed from the off-diagonal Green's functions $g_{12}$ and $g_{21}$. The regular part $\Pi^{\rm R}$ is free from the infrared divergence. In fact, the singular part $\Pi_{14}$ is completely eliminated from $\Pi^{\rm R}$. 
\par
\begin{figure}
\begin{center}
\includegraphics[width=8cm]{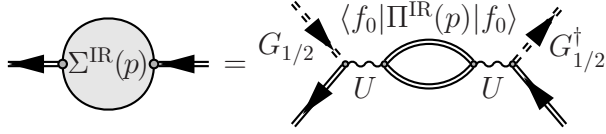}
\end{center}
\caption{Self-energy $\Sigma^{\rm IR}$ used for $\tilde \Sigma$. We take the single-bubble structure for $\Sigma^{\rm IR}$. The wavy line describes the repulsive interaction $U$. The dashed arrow describes the condensate Green's functions $G_{1/2}$ and $G_{1/2}^{\dag}$. }
\label{fig2.fig}
\end{figure}

 
Using $\Pi^{\rm IR}(p)$, we construct $\tilde \Sigma$ in Eq.~(\ref{eq.self2}). We consider the single bubble diagram in Fig.~\ref{fig2.fig}, which provides 
\begin{align} 
{ \Sigma}^{\rm IR} (p) = & - \frac{1}{2} { G}_{1/2} U \langle f_{0} |{ \Pi}^{\rm IR} (p) | f_{0} \rangle U { G}_{1/2}^{\dag}, 
\label{eq4}
\end{align}
where $|f_0\rangle = (0, 1,1, 0)^{\rm T}$. In Eq.~(\ref{eq4}), $G_{1/2} = \sqrt{-n_{0}} (1,1)^{\rm T}$ and $G_{1/2}^{\dag} = \sqrt{-n_{0}} (1,1)$ are the condensate Green's functions.
\par 
\par
\begin{figure}
\begin{center}
\includegraphics[width=8cm]{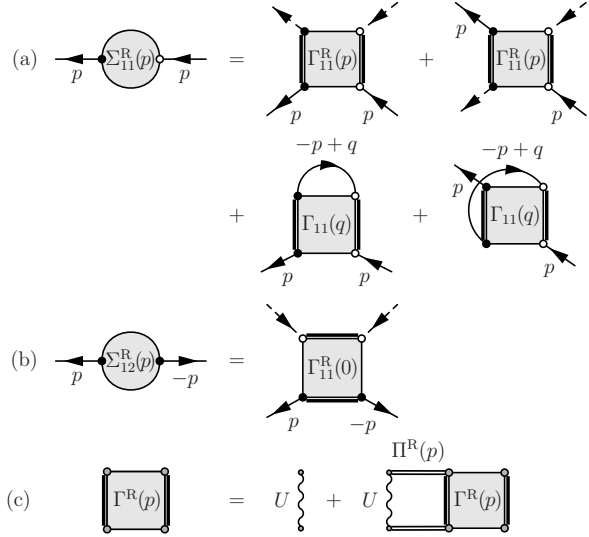}
\end{center}
\caption{Self-energy $\Sigma^{\rm R}$ in the many-body $T$-matrix approximation. (a) Diagonal component $\Sigma_{11}^{\rm R}$. (b) Off-diagonal component $\Sigma_{12}^{\rm R}$. The dashed arrows describe $\sqrt{n_0}$. (c) Bethe-Salpeter equation of the four-point vertex function $\Gamma^{\rm R}$.}
\label{fig3.fig}
\end{figure}

  
We calculate the regular part $\Sigma^{\rm R}(p)$ in Eq.~(\ref{eq.self1}) so as to be free from the infrared divergence. In the MBTA, summing up the diagrams in Fig.~\ref{fig3.fig}, we obtain 
\begin{align}
\Sigma^{\rm R}_{11}(p) = & 2 n_{0} \Gamma^{\rm R}_{11}(p) - 2 T \sum_{q} \Gamma_{11} (q) g_{11} (-p+q), 
\label{Eq243}
\\
\Sigma^{\rm R}_{12}(p) = & n_{0} \Gamma^{\rm R}_{11}(0), 
\label{Eq244}
\end{align} 
where $\Gamma^{\rm R}_{ij} (p)$ is the $(4 \times 4)$-matrix four-point vertex, given by 
\begin{align}
{ \Gamma}^{\rm R} (p) = & \frac{U}{1-U{ \Pi}^{\rm R}(p)}. 
\label{RPABetheSalpeter0} 
\end{align} 
The four-point vertex $\Gamma$ in the second term of Eq.~(\ref{Eq243}) involves the singular part $\Pi^{\rm IR}$, giving the form 
\begin{align}
{ \Gamma}(p) = & \frac{U}{1-U[\Pi^{\rm R}(p)+\Pi^{\rm IR}(p)]}. 
\label{RPABetheSalpeter1} 
\end{align} 
The regular part $\Sigma^{\rm R}$ does not exhibit infrared divergence. The first terms in Eqs.~(\ref{Eq243}) and (\ref{Eq244}) involve $\Pi^{\rm R}$, which are free from the infrared divergence. The second term in Eq.~(\ref{Eq243}) does not exhibit the infrared divergence of $\Pi^{\rm IR}$, after carrying out the summation with respect to the internal momentum $q=({\bf q},\omega_m)$. 
\par 
\par
\begin{figure}
\begin{center}
\includegraphics[width=9cm]{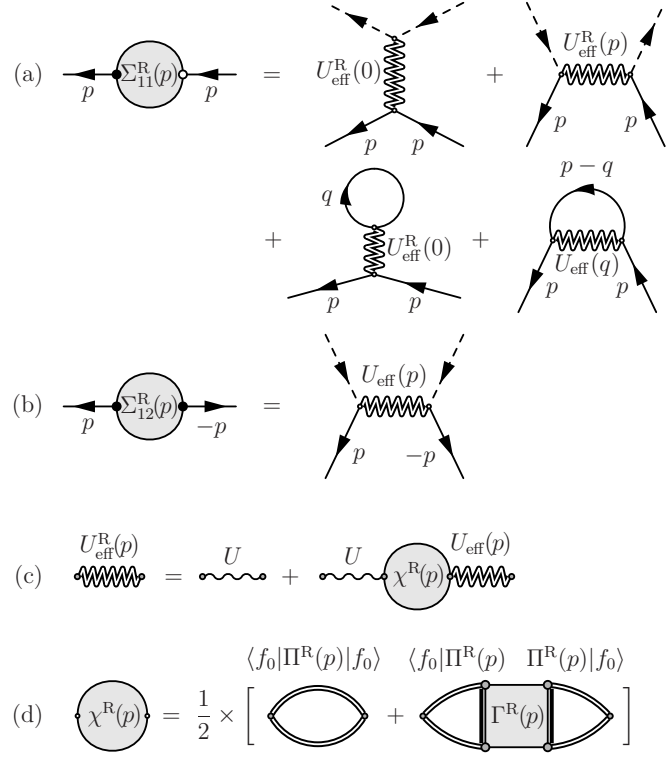}
\end{center}
\caption{Self-energy $\Sigma^{\rm R}$ in the random phase approximation with the vertex correction. (a) Diagonal component $\Sigma_{11}^{\rm R}$. (b) Off-diagonal component $\Sigma_{12}^{\rm R}$. (c) Effective interaction $U_{\rm eff}^{\rm R} (p)$, which involves the density fluctuation effects. (d) The correlation function $\chi^{\rm R}$.}
\label{fig4.fig}
\end{figure}

  
The regular part of the RPA self-energy $\Sigma^{\rm R}$ is also obtained in the same manner. Summing up the diagrams in Fig.~\ref{fig4.fig}, one has 
\begin{align}
\Sigma^{\rm R}_{11}(p) =& (n_{0}+n') U^{\rm R}_{\rm eff}(0) + n_{0} U^{\rm R}_{\rm eff}(p) - T \sum\limits_{q} U_{\rm eff} (q) g_{11} (p-q), 
\label{eq27} 
\\
\Sigma^{\rm R}_{12}(p) = & n_{0} U^{\rm R}_{\rm eff}(p), 
\label{eq28} 
\end{align} 
where $n'=-T \sum_{p} g_{11}(p) e^{i \omega_{n} \delta}$ is the non-condensate density, and 
\begin{align} 
U^{\rm R}_{\rm eff} (p) = & \frac{U}{1 - U \chi^{\rm R}(p) }.
\label{RPABetheSalpeter}
\end{align} 
Here, the correlation function $\chi^{\rm R}$ is given by~\cite{Watabe2013}, 
\begin{align}
\chi^{\rm R} (p) = & \frac{1}{2} \langle f_0 | [ \Pi^{\rm R}(p) + \Pi^{\rm R}(p) \Gamma^{\rm R}(p)\Pi^{\rm R}(p) ] |f_0 \rangle. 
\label{eq10chiR}
\end{align} 
In Eq.~(\ref{eq27}), $U_{\rm eff}$ is also given by Eq.~(\ref{RPABetheSalpeter}), where both $\Pi^{\rm R}$ and $\Gamma^{\rm R}$ in Eq.~(\ref{eq10chiR}) are replaced by $\Pi=\Pi^{\rm R}+\Pi^{\rm IR}$ and $\Gamma$ in Eq.~(\ref{RPABetheSalpeter1}), respectively. As in the MBTA, the infrared singularity in $U_{\rm eff}$ does not remain in the final result $\Sigma_{11}^{\rm R}$ after taking the $q$-summation. 
\par 
\par 
The factors $\tilde G_{\rm L}(p)$ and $\tilde G_{\rm R}(p)$ in Eq.~(\ref{eq.intro5}) are also evaluated in a diagrammatic manner. In this paper, we consider the following two cases as typical examples. Recalling the expression in Eq.~(\ref{eq.intro4}), one may take, as the first example, 
\begin{equation}
\tilde G^{(1)}_{\rm L,\rm R}(p)= {\tilde G}(p).
\label{eq.alpha1}
\end{equation} 
As the second example, we may employ the simple version that involves the self-energy in the Hartree-Fock (HF) approximation, given by
\begin{equation}
\tilde G^{(2)}_{\rm L,\rm R}(p)= {1 \over i\omega_n\sigma_3- \varepsilon_{\bf p} + \mu -\Sigma_{\rm HF}(p)}, 
\label{eq.alpha2}
\end{equation} 
where $\Sigma_{\rm HF}(p)=2U (n_0+n')$ is the HF self-energy. 
\par
These two examples lead to the same infrared singularity in the second term of Eq.~(\ref{eq.intro5}), giving the form 
\begin{eqnarray}
\tilde G_{\rm L}^{(1)}(p){\tilde \Sigma}(p) \tilde G^{(1)}_{\rm R}(p)\Bigr|_{p\to 0} = \frac{n_{0} U^{2} \Pi_{14} (p)}{2 [\Sigma^{\rm R}_{12}(0)]^{2}} 
\begin{pmatrix} 1 & 1 \\ 1 & 1 \end{pmatrix} , 
\label{eq24NN}
\end{eqnarray} 
\begin{eqnarray}
\tilde G_{\rm L}^{(2)}(p){\tilde \Sigma}(p) \tilde G^{(2)}_{\rm R}(p)\Bigr|_{p\to 0} = {2n_{0}U^{2} \Pi_{14} (p) \over [\mu - \Sigma_{\rm HF}(0)]^{2}} 
\begin{pmatrix} 1 & 1 \\ 1 & 1 \end{pmatrix}. 
\label{eq25NN}
\end{eqnarray} 
Both Eqs.~(\ref{eq24NN}) and (\ref{eq25NN}) diverge reflecting the infrared singularity of the correlation function $\Pi_{\rm 14}(p\to 0)$ in Eq.~(\ref{eq10}). Indeed, we have $\Sigma_{12}^{\rm R} (0) \neq 0$, as well as $\mu-\Sigma^{\rm HF} \neq 0$ below $T_{\rm c}$ (at least in the MBTA and the RPA with the vertex correction we are using). As will be shown in the next section, this infrared divergence is a crucial key to reproduce the infrared divergence of the longitudinal response function $\chi_\parallel(p)$, as well as the NN identity. 
\par 
The present approach separately evaluates each term in Eq.~(\ref{eq.intro5}) in a diagrammatic manner, so that one needs to be careful about double counting of many-body corrections. In this regard, we emphasize that this problem is safely avoided in our formalism, because $\Sigma_{\rm a}$ and $\tilde \Sigma$ involve qualitatively different diagrams with respect to the infrared behavior. We briefly note that, although the second term in Eq.~(\ref{eq.intro5}) exhibits the infrared divergence in our approach, it is expected that this contribution is still weaker than the first term in Eq.~(\ref{eq.intro5}). Indeed, for a small ${\bf p}$, the first term in Eq.~(\ref{eq.intro5}) exhibits $\tilde G (0, {\bf p}) \propto {\bf p}^{-2}$. This divergence is stronger than that of the second term in Eq.~(\ref{eq.intro5}), (which is proportional to $\Pi_{14}$ in Eq.~(\ref{eq10})). 
\par 
\section{Longitudinal susceptibility and Nepomnyashchii--Nepomnyashchii identity}\label{SecIII}
\par
\subsection{Longitudinal susceptibility}\label{SecIIIA}
\par 
The longitudinal response function $\chi_{\parallel}$ and the transverse response function $\chi_{\perp}$ are given by~\cite{Weichman1988,Giorgini1992} \begin{align} 
\chi_{\nu} (p) = & \int_{0}^{1/T} d\tau e^{i \omega_{n} \tau} \langle T_{\tau}  a_{\nu \bf p} (\tau) a_{\nu -\bf p} (0) \rangle , 
\label{Eq217} 
\end{align}  
where $T_{\tau}$ denotes a $\tau$-ordering operation, and $\nu \equiv (\parallel, \perp)$.
In Eq.~(\ref{Eq217}), $a_{\parallel {\bf p}}$ and $a_{\perp {\bf p}}$ are longitudinal and transverse operators, respectively. When the BEC order parameter is taken to be real, they are respectively given by 
\begin{align}
a_{\parallel {\bf p}} = \frac{1}{2 } ( a_{\bf p} +a_{- {\bf p}}^{\dag} ) , 
\quad 
a_{\perp {\bf p}} = \frac{1}{2 i } (a_{\bf p} - a_{- {\bf p}}^{\dag} ).
\label{Eq194}
\end{align} 
Equation (\ref{Eq217}) can be also written as 
\begin{align} 
\chi_{\parallel} (p) = - \frac{1}{4} \langle + | G (p) | + \rangle , 
\quad 
\chi_{\perp} (p) = - \frac{1}{4 } \langle - | G (p) | - \rangle , 
\label{chiParallelPerp}
\end{align} 
where $| \pm \rangle \equiv (1, \pm1)^{\rm T}$.
\par 
We here summarize exact properties of these static susceptibilities obtained from the exact Green's function with the self-energy that satisfies the NN identity~\cite{Nepomnyashchii1978}, as well as the Hugenholtz-Pines relation~\cite{Hugenholtz1959}. The transverse susceptibility exhibits the infrared divergence as 
\begin{align}
\chi_{\perp} (0, {\bf p}) \simeq \frac{ n_{0} m }{n |{\bf p}|^{2} }. 
\end{align} 
This indicates the instability of this state against an infinitesimal perturbation in the transverse direction (phase fluctuations) of the BEC order parameter. 
\par   
The static longitudinal susceptibility in the low-momentum region is dominated by the off-diagonal self-energy~\cite{Nepo1983,Dupuis2011LP,Dupuis2011}, given by  
\begin{align}
\chi_{\parallel} (0, {\bf p}) \simeq \frac{1}{4 \Sigma_{12} (0, {\bf p}) }. 
\label{chiparaSiga12}
\end{align} 
Because of the NN identity $\Sigma_{12} (0) = 0$, Eq.~(\ref{chiparaSiga12}) diverges when ${\bf p}=0$. This infrared divergence is, however, weaker than that of the transverse susceptibility, i.e.,~\cite{note1} 
\begin{align}
\chi_{\perp} (0, {\bf p}) \gg \chi_{\parallel} (0, {\bf p}). 
\end{align} 
\par
\begin{figure}
\begin{center}
\includegraphics[width=8cm]{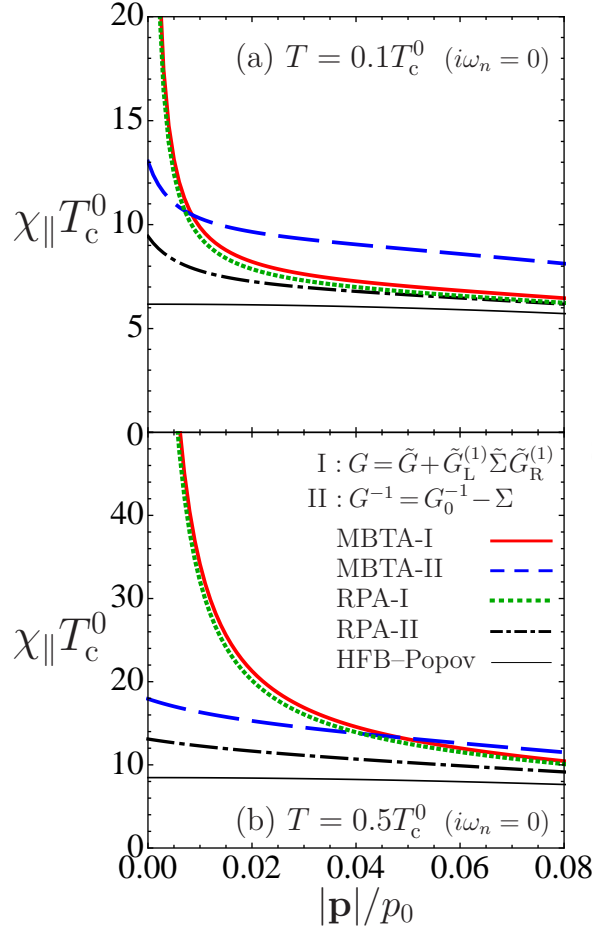}
\end{center}
\caption{(Color online) Static longitudinal susceptibility $\chi_{\parallel} (i\omega_{n} = 0, {\bf p})$. (a) $T = 0.1T_{\rm c}^{0}$. (b) $T = 0.5T_{\rm c}^{0}$. Here, $T_{\rm c}^{0}$ is the critical temperature of an ideal Bose gas. We consider two approximations (the many-body $T$-matrix approximation (MBTA) as well as the random-phase approximation (RPA) with the vertex correction). For each approximation, we apply two different formalisms. One is our formalism in Eq.~(\ref{eq.intro5}) with $\tilde G_{\rm L,R} = \tilde G_{\rm L,R}^{(1)}= \tilde G$, given in Eq.~(\ref{eq.alpha1}). The other is the Green's function obtained from the standard formalism in Eq.~(\ref{eq.intro1}). In self-energies in the standard formalism in Eq.~(\ref{eq.intro1}), we replace $\Gamma_{11}^{\rm R}$ with $\Gamma_{11}$ in the MBTA in Eqs.~(\ref{Eq243}) and (\ref{Eq244}), and also replace $U_{\rm eff}^{\rm R}$ with $U_{\rm eff}$ in the RPA in Eqs.~(\ref{eq27}) and (\ref{eq28}). We set $an^{1/3} = 10^{-2}$ and $p_{\rm c} = 5 p_{0}$, where $p_{0} = \sqrt{2m T_{\rm c}^{0}}$. }
\label{fig5.fig}
\end{figure}

 
The infrared divergence of the longitudinal susceptibility can be correctly described by our approach (Fig.~\ref{fig5.fig}). This result is quite different from the cases of the HFB--Popov approximation, the MBTA, as well as the RPA with the vertex corrections, based on the standard formalism in Eq.~(\ref{eq.intro1})~\cite{Watabe2013,note2}. 
\par 
\par 
In the present case, the longitudinal susceptibility in the limit $p\to0$ behaves as
\begin{eqnarray}
\chi_{\parallel}^{(1)} (p) \simeq - \frac{n_{0} U^{2} \Pi_{14} (p)}{2 [\Sigma^{\rm R}_{12}(0)]^{2}}, 
\label{eq24NNchip}
\end{eqnarray} 
\begin{eqnarray}
\chi_{\parallel}^{(2)} (p) \simeq - {2n_{0}U^{2} \Pi_{14} (p) \over [\mu - \Sigma_{\rm HF}(0)]^{2}}. 
\label{eq25NNchip}
\end{eqnarray} 
Here, $\chi_{\parallel}^{(1)}$ and $\chi_{\parallel}^{(2)}$ are the longitudinal susceptibility in the cases of Eqs.~(\ref{eq.alpha1}) and (\ref{eq.alpha2}), respectively. The second term in Eq.~(\ref{eq.intro5}) provides the infrared divergence of $\chi_{\parallel}$ thanks to the singularity of $\Pi_{14}(p)$. (See also Eqs.~(\ref{eq24NN}) and (\ref{eq25NN}).) This result is consistent with the previous work dealing with the infrared divergence of the longitudinal susceptibility~\cite{Patashinskii1973,Nepo1983,Weichman1988,Giorgini1992,Griffin1993BOOK,Castellani1997,Zwerger2004,Sinner2010,Dupuis2011}. 
\par 
The Bogoliubov approximation fails to reproduce this infrared-divergent longitudinal susceptibility. In this approximation, one obtains~\cite{Nepo1983,Weichman1988} 
\begin{align} 
\chi_{\perp} (0, {\bf p}) \simeq \frac{m}{{\bf p}^{2}}, \quad \chi_{\parallel} (0, {\bf p}) \simeq \frac{1}{4 m c_{0}^{2}}. 
\end{align} 
The origin of the finite longitudinal susceptibility is considered to be decoupling of transverse and longitudinal fluctuations in the mean-field theory~\cite{Weichman1988}. Indeed, the Hamiltonian in the Bogoliubov theory has the form~\cite{Weichman1988} 
\begin{align}
H_{2} = & \sum\limits_{{\bf p} \neq {\bf 0}} \biggl [ F_{0} a_{\perp {\bf p}} a_{\perp - {\bf p}} + F_{2} a_{\parallel {\bf p}} a_{\parallel - {\bf p}} + \frac{1}{2} F_{1} \biggr ], 
\end{align} 
where $F_{j} = \varepsilon_{\bf p} + j U n_{0}$. Anharmonic effects of fluctuations were pointed out to be important to obtain the infrared-divergent longitudinal susceptibility~\cite{Weichman1988}. 
\par 
\subsection{Nepomnyashchii--Nepomnyashchii identity}\label{SecIIIB}
\par 
Our Green's function also satisfies the NN identity~\cite{Nepomnyashchii1975,Nepomnyashchii1978}, as well as the Hugenholtz-Pines relation~\cite{Hugenholtz1959}. These two required conditions are conveniently summarized as
\begin{align}
{ \Sigma} (0) = \mu. 
\label{Eq214}
\end{align} 
Given the hybrid version of the Green's function in Eq.~(\ref{eq.intro5}) in the standard expression in Eq.~(\ref{eq.intro1}), one finds that the self-energy in Eq. (\ref{eq.intro1}) has the form 
\begin{align}
{ \Sigma} (p) = & G_{0}^{-1} (p) - \frac{1}{1 + \tilde G^{-1} (p) \tilde G_{\rm L}(p) \tilde \Sigma (p) \tilde G_{\rm R} (p)} \tilde G^{-1} (p) .
\label{eq12}
\end{align} 
In Eq.~(\ref{eq12}), the factor $\tilde G_{\rm L} \tilde {\Sigma} \tilde G_{\rm R}$ involves the infrared divergence (as seen in Eqs.~(\ref{eq24NN}) and (\ref{eq25NN})), whereas $\tilde G^{-1}(p\to 0)$ safely converges in the MBTA as well as the RPA with the vertex correction. The second term thus vanishes when $p\to0$, leading to Eq.~(\ref{Eq214}) through the relation ${ \Sigma} (0) = { G}_{0}^{-1} (0)=\mu$. 
\par 
The present approach in Eq.~(\ref{eq.intro5}) may be viewed as an extension of the Popov's hydrodynamic approach at $T=0$~\cite{Popov1972,Popov1979,Popov1983BOOK} to the finite temperature region. Far below $T_{\rm c}$ in the weak coupling regime where the non-condensate density is negligible, one may retain the regular self-energy $\Sigma^{\rm R}$ to the lowest order, giving the form 
\begin{align}
\Sigma_{11}^{\rm R} = 2 U n_{0}, \quad \Sigma_{12}^{\rm R} = U n_{0}. 
\label{BogoParam}
\end{align} 
When we apply Eq.~(\ref{BogoParam}) to $\tilde G_{\rm L,R}$ in Eq.~(\ref{eq.intro5}), one has
\begin{align}
\tilde G_{\rm L,R} (p) = g (p). 
\label{eq42}
\end{align}
For $\tilde \Sigma$, we use Eq.~(\ref{eq4}). In addition, we assume the hydrodynamic regime $|{\bf p}| \ll \sqrt{2} m c_{0}$ (where $c_{0}$ is the Bogoliubov sound speed). Then, the Green's function (\ref{eq.intro5}) is reduced to 
\begin{align}
G (p) = - \frac{ m c_{0}^{2}}{\omega_{n}^{2} + c_{0}^{2} {\bf p}^{2}} \begin{pmatrix} 1 & -1 \\ -1 & 1 \end{pmatrix} + \frac{\Pi_{14} (p)}{2 n_{0}} \begin{pmatrix} 1 & 1 \\ 1 & 1 \end{pmatrix} . 
\label{hydroG1}
\end{align} 
Equation (\ref{hydroG1}) equals the Green's function in the Popov's hydrodynamic theory (which is explained in Appendix~\ref{AppendixB}).  
\par 
We note that the phase (transverse) fluctuation affects the amplitude (longitudinal) fluctuation, and leads to the infrared divergence of the longitudinal susceptibility~\cite{Podolsky2011}. 
Indeed, according to the Popov's hydrodynamic theory, the second term in Eq.~(\ref{hydroG1}) providing the infrared divergence of $\chi_{\parallel}$ originates from the convolution of the phase-phase correlation. (See also Appendix~\ref{AppendixB}.)
\par 
We also note that Eq.~(\ref{hydroG1}) has the same structure as the exact Green's function in the low-energy and low-momentum limit obtained by Nepomnyashchii and Nepomnyashchii~\cite{Nepomnyashchii1978}, 
\begin{align} 
G (p) = & \frac{n_{0} m c^{2}}{n} \frac{ 1 }{ \omega_{}^{2} - c^{2} |{\bf p}|^{2} } \begin{pmatrix} 1 & -1 \\ -1 & 1 \end{pmatrix} - \frac{1}{4 \Sigma_{12} (p) } \begin{pmatrix} 1 & 1 \\ 1 & 1 \end{pmatrix}, 
\label{Eq182} 
\end{align} 
where $c$ is the macroscopic sound velocity determined from the compressibility. 
\par
\begin{figure}
\begin{center}
\includegraphics[width=8.5cm]{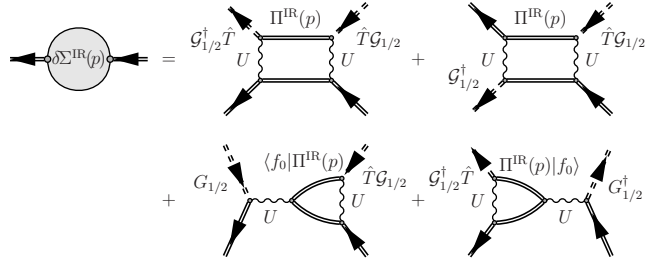}
\end{center}
\caption{Self-energy $\delta \Sigma^{\rm IR}$. }
\label{fig6.fig}
\end{figure}

 
The single bubble diagram giving Eq.~(\ref{eq4}) is the primitive many-body correction to the self-energy $\tilde \Sigma$ to reproduce the NN identity. Any other $p$-dependent second-order corrections do not contribute to $\tilde \Sigma$. To explicitly see this, we conveniently write $\tilde \Sigma$ in the form 
\begin{align}
\tilde \Sigma (p) = & \Sigma^{\rm IR} (p) + \delta \Sigma_{}^{\rm IR} (p), 
\label{eq3dash2nd} 
\end{align} 
where $\Sigma^{\rm IR} (p)$ is given in Eq.~(\ref{eq4}), and $\delta \Sigma^{\rm IR} (p)$ is diagrammatically described as Fig.~\ref{fig6.fig}, which gives
\begin{align} 
\delta \Sigma^{\rm IR}  (p) = & - {\mathcal G}_{1/2}^{\dag} \hat T U { \Pi}^{\rm IR} (p) U \hat T {\mathcal G}_{1/2} 
\nonumber 
\\ & 
- {\mathcal G}_{1/2}^{\dag} U { \Pi}^{\rm IR} (p) U \hat T {\mathcal G}_{1/2}  
\nonumber 
\\ & 
- G_{1/2} U \langle f_{0} | { \Pi}^{\rm IR} (p) U \hat T {\mathcal G}_{1/2}
\nonumber 
\\ & 
- {\mathcal G}_{1/2}^{\dag} \hat T U { \Pi}^{\rm IR} (p) | f_{0} \rangle U G_{1/2}^{\dag}.
\label{SigmatildeIR}
\end{align} 
Here, we have introduced matrix condensate Green's functions ${\mathcal G}_{1/2} = \sqrt{-n_{0}} \hat \eta_{g}$ and ${\mathcal G}_{1/2}^{\dag} = \sqrt{-n_{0}} \hat \eta_{g}^{\dag}$. The matrices $\hat T$ and $\hat \eta_{g}^{}$ are given by, respectively,  
\begin{align}
\hat T= \begin{pmatrix} 1 & 0 & 0 & 0 \\ 0 & 0 & 1 & 0 \\ 0 & 1 & 0 & 0 \\ 0 & 0 & 0 & 1 \end{pmatrix}, 
\qquad 
\hat \eta_{g}^{} = \begin{pmatrix} 1 & 0 \\ 1 & 0 \\ 0 & 1 \\ 0 & 1 \end{pmatrix} . 
\end{align} 
Using $\hat C \hat T \hat \eta_{g} = \hat C \hat \eta_{g} = 0$, we find that $\delta { \Sigma}^{\rm IR} (p) = 0$, as expected.
\par  
\par
\begin{figure}
\begin{center}
\includegraphics[width=9cm]{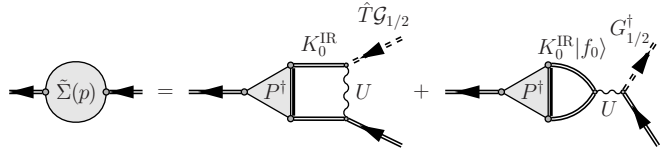}
\end{center}
\caption{Self-energy $\tilde \Sigma$ with the vertex correction $P^{\dag}$. }
\label{fig7.fig}
\end{figure}

We point out that the NN identity still obtains even if one includes vertex corrections to the single bubble contribution in Eq.~(\ref{eq4}). Indeed, considering the self-energy corrections given in Fig.~\ref{fig7.fig}, we have 
\begin{align}
\tilde \Sigma (p) = & - T \sum\limits_{q} P^{\dag} (q;p)  K_{0}^{\rm IR} (q;p) U \frac{1}{\sqrt{-1}} \hat T {\mathcal G}_{1/2} 
\label{SigmaIRvc}
\\
& 
- \frac{1}{2 } T \sum\limits_{q} P^{\dag} (q;p) K_{0}^{\rm IR} (q;p) |f_0 \rangle U \frac{1}{\sqrt{-1}} G_{1/2}^{\dag},  
\nonumber 
\end{align} 
where $P^{\dag} (q;p)$ is a $(2\times 4)$-matrix three-point vertex, and ${ K}_{0}^{\rm IR} (q;p) \equiv K^{0}_{1212} (q;p) \hat C$ with $K^{0}_{1212} (q;p) \equiv g_{12}(p+q) g_{12} (-q)$. Although $K_{0}^{\rm IR}$ provides the singular part $\Pi^{\rm IR}$ in Eq.~(\ref{eq.secIII13}), the first term in Eq.~(\ref{SigmaIRvc}) actually does not exhibit the infrared divergence, because $\hat C \hat T \hat \eta_{g} = 0$.
\par 
\par 
To examine the infrared behavior of the second term in (\ref{SigmaIRvc}), 
it is convenient to use the Ward-Takahashi identity with respect to the three-point vertex $P^{\dag}$ and the off-diagonal self-energy $\Sigma_{12} (0)$ in the limit $p\to 0$~\cite{Nepomnyashchii1978}, giving the form 
\begin{align} 
P^{\dag} (0;0) = & 2 \frac{\Sigma_{12}(0)}{\sqrt{n_{0}}} \eta^{\dag} + n_{0}^{3/2} \frac{\partial}{\partial n_{0}} \left ( \frac{\Sigma_{12}(0)}{n_{0}} \right ) \eta_{\rm a}^{\dag} , 
\label{2-3relation}
\end{align} 
where 
\begin{align} 
\eta^{\dag} = \begin{pmatrix} 1 & 1 &  1 & 0 \\ 0 & 1 & 1 & 1 \end{pmatrix} , 
\quad 
\eta_{\rm a}^{\dag} = \begin{pmatrix} 1 & 1 &  1 & 1 \\ 1 & 1 & 1 & 1 \end{pmatrix} . 
\end{align} 
(For the derivation, see Appendix~\ref{AppendixC1}.) The contribution $P^{\dag} (0;0)$ can be clearly extracted from Eq.~(\ref{SigmaIRvc}), if we apply the second mean value theorem for integrals~\cite{Iyanaga1977}. Using this theorem, the second term in (\ref{SigmaIRvc}) can be divided into two terms: the term involving the infrared divergence and the term which is finite in the limit $p\to 0$. The former involves $P^{\dag} (0 ;0) \Pi_{14}^{\Lambda} \hat C$, where $\Pi_{14}^{\Lambda} = - T \sum_{{\bf q}=0}^{\Lambda} K_{1212}^{0}(i \omega_{n} = 0,{\bf q};0)$, and $\Lambda$ is a cutoff determined from the mean value theorem. Then, using the relation $\eta^{\dag} \hat C | f_{0} \rangle  = 2 | + \rangle$, one finds
\begin{align}
\tilde { \Sigma} (0) = & 2 U \Pi_{14}^{\Lambda} \Sigma_{12} (0) \begin{pmatrix} 1 & 1 \\ 1 & 1 \end{pmatrix} + \delta \Sigma, 
\label{eq19}
\end{align} 
where $\delta \Sigma_{} $ is the non-singular part of $\tilde \Sigma $. When we evaluate the self-energy $\Sigma = G_{0}^{-1} - G^{-1}$ using Eq.~(\ref{eq19}) together with $\Pi_{14}^{\Lambda}(0) = \infty$, we reach the expected result $\Sigma (0) = \mu$. 
\par 
Nepomnyashchii and Nepomnyashchii derived the NN identity in a similar manner~\cite{Nepomnyashchii1978}. The NN identity is ascribed to the infrared divergence in the bubble structure self-energy with the vertex correction. 
Diagrams providing required infrared behaviors of $\chi_{\parallel}(0)$ as well as $\Sigma_{12}(0)$ are common between our formalism and the exact results studied by Nepomnyashchii and Nepomnyashchii~\cite{Nepomnyashchii1978}. The original derivation of the NN identity as well as the relation to the phase fluctuation are summarized in Appendices~\ref{AppendixC2} and~\ref{AppendixC3}, respectively.
\par
\section{Condensate fraction}\label{SecIV}
\par 
\begin{figure}
\begin{center}
\includegraphics[width=8cm]{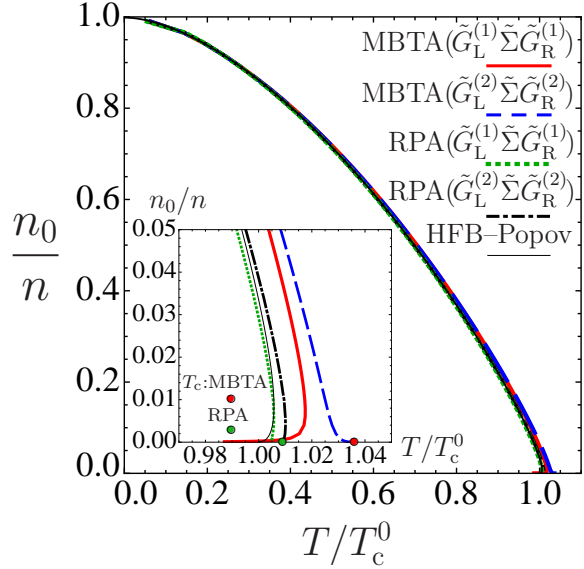}
\end{center}
\caption{(Color online) Condensate fraction $n_0$ calculated in our formalism (\ref{eq.intro5}). We separately examine the many-body $T$-matrix approximation (MBTA) as well as the random-phase approximation (RPA) with the vertex correction. We employ approximations $\tilde G_{\rm L,R}^{(1)}$ in (\ref{eq.alpha1}) as well as $\tilde G_{\rm L,R}^{(2)}$ in (\ref{eq.alpha2}) on $\tilde G_{\rm L,R}$. The inset is the condensate density $n_0$ magnified near $T_{\rm c}$. The points plotted in the inset are the critical temperature $T_{\rm c}$ in each approximation evaluated from the region above $T_{\rm c}$. This temperature $T_{\rm c}$ is common between two cases $\tilde G_{\rm L}^{(1)} \tilde \Sigma \tilde G_{\rm R}^{(1)}$ and $\tilde G_{\rm L}^{(2)} \tilde \Sigma \tilde G_{\rm R}^{(2)}$, because $\tilde \Sigma = 0$ at $T_{\rm c}$. We set $an^{1/3} = 10^{-2}$ and $p_{\rm c} = 5 p_{0}$. 
In the Hartree--Fock--Bogoliubov (HFB)--Popov approximation, we apply the interaction strength $U_0=4\pi a/m$, instead of $U$. 
}
\label{fig8.fig}
\end{figure} 
\par 
Figure \ref{fig8.fig} shows the condensate fraction $n_0$ in the BEC phase, calculated from the equation,
\begin{equation}
n_0=n-n'=n+T \sum_{p} G_{11} (p) e^{i\omega_{n} \delta}, 
\end{equation} 
where the second term $n'$ is the non-condensate density. Many-body corrections to $T_{\rm c}$ are dominated by the first term in Eq.~(\ref{eq.intro5}), when $T_{\rm c}$ is determined from the theory above $T_{\rm c}$. Indeed, both $\delta \tilde G^{(1)} \equiv \tilde G_{\rm L}^{(1)} \tilde \Sigma \tilde G_{\rm R}^{(1)}$ and $\delta \tilde G^{(2)} \equiv \tilde G_{\rm L}^{(2)} \tilde \Sigma \tilde G_{\rm R}^{(2)}$ are absent in the normal state above $T_{\rm c}$. As a result, the enhancement of $T_{\rm c}$ in each case of the MBTA and the RPA with the vertex correction is the same as our previous results based on the standard Green's function formalism in Eq.~(\ref{eq.intro1})~\cite{Watabe2013}. Given the shift of $T_{\rm c}$ as
\begin{align}
\frac{ T_{\rm c}-T_{\rm c}^{0} }{ T_{\rm c}^{0} } = c_{1} an^{1/3}, 
\end{align}
one finds $c_{1} \simeq 3.9$ in the MBTA and $c_{1} \simeq 1.1$ in the RPA with the vertex correction~\cite{Watabe2013} (where $T_{\rm c}^{0}$ is the phase transition temperature in an ideal Bose gas). 
The RPA result is close to the Monte-Carlo result $c_1\simeq1.3$~\cite{Arnold2001,Kashurnikov2001,Nho2004}, whereas the MBTA overestimates the coefficient $c_{1}$. 
\par 
When $T_{\rm c}$ is evaluated by the theory below $T_{\rm c}$, $\delta \tilde G^{(1)}$ and $\delta \tilde G^{(2)}$ give different results. In the case of $\delta \tilde G^{(2)}$, the contribution of $\delta G_{}^{(2)}$ smoothly vanishes in the limit $n_{0} \rightarrow 0$, so that the value of $T_{\rm c}$ coincides with that evaluated from the region above $T_{\rm c}$. The order of the phase transition is also the same as the result based on the standard Green's function formalism in Eq.~(\ref{eq.intro1})~\cite{Watabe2013}. When the self-energy in the first term of Eq.~(\ref{eq.intro5}) is treated within the RPA with the vertex correction, the weak first-order phase transition is obtained. In the case of the MBTA, one obtains the second-order phase transition. (See the inset of Fig.~\ref{fig8.fig}.)
\par 
In the case of $\delta \tilde G^{(1)}$, on the other hand, $T_{\rm c}$ determined by the temperature in the limit $n_0\to +0$ does not coincide with $T_{\rm c}$ determined by the theory above $T_{\rm c}$. In addition, in both the cases of the MBTA and the RPA with the vertex correction, the condensate fraction $n_0$ exhibits a remarkable reentrant behavior near $T_{\rm c}$. (See the inset in Fig.~\ref{fig8.fig}.) These are ascribed to large contribution of Eq.~(\ref{eq24NN}). Given the regular part of the off-diagonal self-energy 
\begin{equation}
\Sigma_{12}^{\rm R} (0)=n_0 V_{\rm eff}, 
\end{equation}
(where $V_{\rm eff} = \Gamma_{11}^{\rm R}(0)$ in the MBTA, and $V_{\rm eff} = U_{\rm eff}^{\rm R}(0)$ in the RPA), we find that in the limit $p\to 0$, Eq.~(\ref{eq24NN}) is reduced to 
\begin{align} 
\tilde G_{\rm L}^{(1)} (p) \tilde \Sigma (p) \tilde G_{\rm R}^{(1)} (p) \simeq \frac{ U^{2} \Pi_{14} (p)}{2 n_{0} V_{\rm eff}^{2}} \begin{pmatrix} 1 & 1 \\ 1 & 1 \end{pmatrix} . 
\label{eq29NN}
\end{align} 
Equation (\ref{eq29NN}) becomes very large, when $n_{0}\to +0$~\cite{note3}. Thus, the temperature has to decrease near $T_{\rm c}$, so as to satisfy the number equation $n = n_{0} + n'$. This leads to the reentrant behavior of $n_0$ seen in the inset of Fig.~\ref{fig8.fig}. 
In addition, while Eq.~(\ref{eq29NN}) is very large in the limit $n_{0} \to + 0$, it is absent in the normal state above $T_{\rm c}$ because $\Pi_{14} = 0$. One thus obtains the discrepancy of the critical temperature between formalisms above and below $T_{\rm c}$. 
\par 
The reentrant behavior of $n_0$ is more remarkable in the MBTA than in the RPA. In the former MBTA, the effective interaction $V_{\rm eff}$ vanishes at $T_{\rm c}$~\cite{Shi1998,Watabe2013}, because 
\begin{align}
\Gamma_{11} (p) = \Gamma_{11}^{\rm R} (p) = \frac{U}{1 - U \Pi_{11} (p)} \quad (T = T_{\rm c}), 
\end{align}
where $\Pi_{11} (0) = \infty$. On the other hand, in the latter RPA with the vertex correction~\cite{Watabe2013}, one finds $V_{\rm eff} = U/2$ at $T_{\rm c}$, because
\begin{align}
U_{\rm eff} (p) = U_{\rm eff}^{\rm R} (p) = \frac{U [ 1- U \Pi_{22} (p) ]}{1 - 2 U \Pi_{22} (p)} \quad (T = T_{\rm c}), 
\end{align}
and $\Pi_{22} (0) = \infty$. As a result, Eq.~(\ref{eq29NN}) is larger in the MBTA than in the RPA with the vertex correction. 
\par 
Although the present approach can correctly describe the infrared behavior of the longitudinal response function, as well as the NN identity, the above results indicate that it still has room for improvement in the fluctuation region near $T_{\rm c}$. In this region, strong fluctuations dominate over the phase transition behavior~\cite{Capogrosso-Sansone2010,Kashurnikov2001,Arnold2001A,Arnold2001B,Prokof'ev2001,Prokof'ev2004}. 
\par
\section{Summary}\label{secIX}
\par  
We have presented a Green's function formalism, which can correctly describe two required conditions for any consistent theory of Bose--Einstein condensates (BECs): (i) the infrared divergence of the longitudinal susceptibility in the low-energy and low-momentum limit, as well as (ii) the Nepomnyashchii--Nepomnyashchii (NN) identity, which states the vanishing off-diagonal self-energy in the same limit. These conditions cannot be satisfied in the Bogoliubov mean-field theory, the many-body $T$-matrix theory (MBTA), as well as the random-phase approximation (RPA) with the vertex correction.
\par 
Our key idea is to divide the irreducible self-energy contribution into the singular and non-singular parts with respect to the infrared divergence. These self-energies are separately included in the Green's function so as to satisfy various conditions that are required for any consistent theory of BECs. In this paper, we treated the non-singular self-energy as the ordinary self-energy correction in the Green's function. On the other hand, we dealt with the singular self-energy to the first-order. The resulting Green's function consists of two terms, which is similar to the Green's function in the Popov's hydrodynamic theory~\cite{Popov1972,Popov1979,Popov1983BOOK}.
\par
The singular component mentioned above enables us to correctly describe the infrared divergence of the longitudinal susceptibility, the Hugenholtz-Pines relation, as well as the NN identity. In addition, we showed that the non-singular part of the self-energy provides the enhancement of the BEC phase transition temperature $T_{\rm c}$ (which has been predicted by various methods). The value of the enhancement depends on to what extent we take into account many-body corrections in the non-singular self-energy. 
\par 
The present approach can describe various required conditions that are not satisfied in the previous theories, such the Bogoliubov-type mean-field theory, the MBTA, as well as the RPA with the vertex correction. On the other hand, it still has room for improvement in considering the region near $T_{\rm c}$. When the non-singular self-energy is treated within the MBTA, the expected second-order phase transition may be obtained, whereas the enhancement of $T_{\rm c}$ is overestimated compared with the Monte-Carlo simulation result. When the non-singular part is calculated within the RPA with the vertex correction, the enhancement of $T_{\rm c}$ is close to the Monte-Carlo result compared with the MBTA, whereas it incorrectly gives the first-order phase transition. The further improvement of the present approach to overcome this problem is a remaining issue. 
\par 
\acknowledgements 
We thank M. Ueda for valuable discussions and comments. We also thank A. Leggett and Y. Takada for discussions. S.W. was supported by JSPS KAKENHI Grant Number (249416). Y.O. was supported by Grant-in-Aid for Scientific research from MEXT in Japan (25400418, 25105511, 23500056).
\par 
\appendix
\section{Polarization Functions}\label{AppendixA}
\par 
\subsection{list of $\Pi$}\label{AppendixA1}
\par 
The polarization functions used in this paper are summarized as follows: 
\begin{widetext}
\begin{align} 
\Pi_{11} (p) = & 
- \sum\limits_{\bf q} 
\frac{1}{2} \left [ (E_{{\bf p}+{\bf q}} - E_{\bf q}) \left ( 1 - \frac{\xi_{{\bf p}+{\bf q}}\xi_{\bf q}}{E_{{\bf p}+{\bf q}}E_{\bf q}} \right ) + i \omega_{n} \left ( \frac{\xi_{{\bf p}+{\bf q}}}{E_{{\bf p}+{\bf q}}} - \frac{\xi_{\bf q}}{E_{\bf q}}  \right ) \right ]
\frac{n_{{\bf p}+{\bf q}} - n_{\bf q}}{ \omega_{n}^{2} + (E_{{\bf p}+{\bf q}} - E_{\bf q})^{2}}
\nonumber 
\\
& \quad
- \sum\limits_{\bf q} 
\frac{1}{2} \left [ (E_{{\bf p}+{\bf q}} + E_{\bf q}) \left ( 1 + \frac{\xi_{{\bf p}+{\bf q}}\xi_{\bf q}}{E_{{\bf p}+{\bf q}}E_{\bf q}} \right ) 
+ i \omega_{n} \left ( \frac{\xi_{{\bf p}+{\bf q}}}{E_{{\bf p}+{\bf q}}} + \frac{\xi_{\bf q}}{E_{\bf q}}  \right ) \right ]
\frac{1 + n_{{\bf p}+{\bf q}} + n_{\bf q}}{ \omega_{n}^{2} + (E_{{\bf p}+{\bf q}} + E_{\bf q})^{2}} , 
\label{gapless Hartree-Fock-BogoliubovPi11fp}
\\
\Pi_{12} (p) = & 
- \sum\limits_{\bf q} 
\frac{1}{2} \Delta \left [ \frac{\xi_{{\bf p}+{\bf q}}}{E_{{\bf p}+{\bf q}}E_{\bf q}} (E_{{\bf p}+{\bf q}} - E_{\bf q}) + \frac{ i \omega_{n} }{E_{\bf q}} \right ]
\frac{n_{{\bf p}+{\bf q}} - n_{\bf q}}{\omega_{n}^{2} + (E_{{\bf p}+{\bf q}} - E_{\bf q})^{2}}
\nonumber 
\\
& \quad
+ \sum\limits_{\bf q} 
\frac{1}{2} \Delta \left [ \frac{\xi_{{\bf p}+{\bf q}}}{E_{{\bf p}+{\bf q}}E_{\bf p}} (E_{{\bf p}+{\bf q}} + E_{\bf q}) + \frac{ i \omega_{n} }{E_{\bf q}} \right ]
\frac{1 + n_{{\bf p}+{\bf q}} + n_{\bf q}}{ \omega_{n}^{2} + (E_{{\bf p}+{\bf q}} + E_{\bf q})^{2}} , 
\label{gapless Hartree-Fock-BogoliubovPi12fp}
\\
\Pi_{14} (p) = & 
\sum\limits_{\bf q} 
\frac{1}{2} \frac{\Delta^{2}}{E_{{\bf p}+{\bf q}}E_{\bf q}}
\left [ 
(E_{{\bf p}+{\bf q}} - E_{\bf q})
\frac{n_{{\bf p}+{\bf q}} - n_{\bf q}}{ \omega_{n}^{2} + (E_{{\bf p}+{\bf q}} - E_{\bf q})^{2}}
- (E_{{\bf p}+{\bf q}} + E_{\bf q})
\frac{1 + n_{{\bf p}+{\bf q}} + n_{\bf q}}{ \omega_{n}^{2} + (E_{{\bf p}+{\bf q}} + E_{\bf q})^{2}}
\right ] , 
\label{gapless Hartree-Fock-BogoliubovPi14fp}
\\
\Pi_{22} (p) = & 
\sum\limits_{\bf q} 
\frac{1}{2} \left [ (E_{{\bf p}+{\bf q}} - E_{\bf q}) \left ( 1 + \frac{\xi_{{\bf p}+{\bf q}}\xi_{\bf q}}{E_{{\bf p}+{\bf q}}E_{\bf q}} \right ) 
+ i \omega_{n} \left ( \frac{\xi_{{\bf p}+{\bf q}}}{E_{{\bf p}+{\bf q}}} + \frac{\xi_{\bf q}}{E_{\bf q}}  \right ) \right ]
\frac{n_{{\bf p}+{\bf q}} - n_{\bf q}}{ \omega_{n}^{2} + (E_{{\bf p}+{\bf q}} - E_{\bf q})^{2}}
\nonumber 
\\
& \quad
+
\sum\limits_{\bf q} 
\frac{1}{2} \left [ (E_{{\bf p}+{\bf q}} + E_{\bf q}) \left ( 1 - \frac{\xi_{{\bf p}+{\bf q}}\xi_{\bf q}}{E_{{\bf p}+{\bf q}}E_{\bf q}} \right ) 
+ i \omega_{n} \left ( \frac{\xi_{{\bf p}+{\bf q}}}{E_{{\bf p}+{\bf q}}} - \frac{\xi_{\bf q}}{E_{\bf q}}  \right ) \right ]
\frac{1 + n_{{\bf p}+{\bf q}} + n_{\bf q}}{ \omega_{n}^{2} + (E_{{\bf p}+{\bf q}} + E_{\bf q})^{2}} , 
\label{gapless Hartree-Fock-BogoliubovPi22fp} 
\end{align}
\end{widetext}
where $\xi_{\bf p} \equiv \varepsilon_{\bf p} + \Delta$, $\Delta \equiv U n_{0}$, $E_{\bf p} \equiv \sqrt{\varepsilon_{\bf p} (\varepsilon_{\bf p} + 2 \Delta )}$, and $n_{\bf p}$ is the Bose distribution function $n_{\bf p} \equiv 1/ ( e^{\beta E_{\bf p}} - 1)$ with $\beta = 1/T$. 
\par 
\subsection{Infrared behaviors of $\Pi_{14}$}\label{AppendixA2} 
\par 
We discuss the infrared properties of the polarization function $\Pi_{14}$ for the system dimensionality $d = 3$. We are going to derive the relation 
\begin{align}
\Pi_{14} (p) \propto 
\left \{
\begin{array}{ll} 
\ln (c_{0}^{2} |{\bf p}|^{2} - \omega^{2} ) & \qquad (T = 0) 
\\ 1/ |{\bf p}| & \qquad (T \neq 0) . 
\end{array} 
\right .
\label{pi14irT0}
\end{align} 
\par 
For simplicity, we use the dimensionless quantities. We scale the energy by the critical temperature of an ideal Bose gas $T_{\rm c}^{0}$. In the dimensionless formula, we use $\tilde E_{\bf p} = E_{\bf p}/T_{\rm c}^{0}$, $\tilde \Delta = \Delta/T_{\rm c}^{0}$, $\tilde \Pi_{14} (q) = \Pi_{14} (q) T_{\rm c}^{0}$, and $\tilde \varepsilon_{\bf p} = \varepsilon_{\bf p}/T_{\rm c}^{0} = \tilde p^{2}$. We wrote the modulus of the momentum in the dimensionless form as $\tilde p = |\tilde {\bf p}|$. In the following, we omit the tilde for simplicity.
\par 
After lengthy calculation, we reduce the polarization function $\Pi_{14}$ as 
\begin{align} 
\Pi_{14} (p) = - A_3 2 \pi \int_{0}^{{p}_{\rm c}} d   q   q^2 \frac{  \Delta^2}{2   E_{\bf q} } \Xi_{14} g_{\bf q} , 
\label{tildePi14}
\end{align} 
where $g_{\bf q} \equiv \coth \left (  \beta E_{\bf q} / 2 \right )$. Here, $\Xi_{14}$ is given by 
\begin{align}
\Xi_{14} = {\rm Re}\left [ \frac{1}{2   p   q R } \ln{ \frac{ (   P_{+} +   \Delta - R ) (  P_{-} +   \Delta + R) }{ (   P_{-} +   \Delta - R ) (   P_{+} +   \Delta + R ) } } \right ] ,  
\end{align} 
where $P_{\pm} = ( p \pm q)^2$, $R = \sqrt{A^2 + \Delta^2}$ and $A = i \omega_n - E_{\bf q}$. The coefficient $A_{3}$ is given by $A_{3} = 1/[\pi^{3/2} \zeta (3/2)]$, and $\zeta $ is the Riemann zeta function. 
\par 
For the small $q$ and $i\omega_{n}$, we have $A \simeq i \omega_{n} - c_{0} q$ and $R \simeq \Delta + c_{0}^{-2} (i \omega_{n} - c_{0} q)^{2}$. In this case, the main contribution of $\Xi_{14}$ reads as 
\begin{align}
\Xi_{14} \simeq & \frac{1}{c_{0}^{2} p q} {\rm Re} \left [ \ln \left ( \frac{A_{+}   q + B}{A_{-}   q + B} \right ) \right ], 
\end{align}
where $A_{\pm} = 2 \left ( \pm p + i \omega_{n} / c_{0} \right )$, and $B = p^{2} - \left ( i \omega_{n} / c_{0} \right )^{2}$. 
\par 
At $T = 0$, we replace $i \omega_{n}$ with $\omega$, and take $g_{\bf q} = 1$. We have 
\begin{align}
\Pi_{14} \simeq & - A_{3} \frac{\pi c_{0}}{4  p} {\rm Re} [  F(  p_{\rm c}) -   F (0) ] , 
\end{align} 
where 
\begin{align}
F (q) = & q \ln \left ( \frac{B + A_{+}   q}{B + A_{-}  q} \right ) + \sum\limits_{j = \pm} \frac{j B}{A_{j}} \ln (B + A_{j} q) .
\end{align} 
The main contribution originates from $F (0)$, and we end with 
\begin{align}
\Pi_{14} \simeq & A_{3} \frac{\pi c_{0}}{4}  \ln \left ( p^{2} - \frac{  \omega^{2}}{c_{0}^{2}} \right ).
\end{align} 
This leads to (\ref{pi14irT0}) for $T=0$. 
\par 
At $T \neq 0$, we take $\omega_{n} = 0$ and $g_{\bf q} = 2   T / c_{0} q$. In this case, we have 
\begin{align}
\Pi_{14} \simeq &  - A_{3} \frac{\pi   T }{2   p} {\rm Re}[ F (  p_{\rm c}) ] , 
\end{align} 
where  
\begin{align}
F ( q ) = & - {\rm Li}_{2} \left ( 4 q / A_{-} \right ) + {\rm Li}_{2} \left ( 4 q / A_{+} \right ) . 
\end{align} 
Here, ${\rm Li}_{n}(z)$ is the polylogarithm (Jonqui\`ere's function). We also used $F (0) = 0$. For large $  p_{\rm c}$, ${\rm Re} [ F (  p_{\rm c}) ] \simeq \pi^{2} / 2  -   p /   p_{\rm c}$ holds. We thus end with 
\begin{align} 
\Pi_{14} \simeq & - A_{3} \frac{\pi^{3}   T}{4} \frac{1}{  p}. 
\label{AppBlast}
\end{align} 
This leads to (\ref{pi14irT0}) for $T\neq 0$. 
\par 
\section{Popov's hydrodynamic theory}\label{AppendixB}
\par 
We derive the single-particle Green's function in the Popov's hydrodynamic theory. We suppose that the system is (a) in the weak coupling regime, (b) at $T \simeq 0$, as well as (c) in the hydrodynamic regime $|{\bf p}| \ll \sqrt{2} m c$. 
\par 
In the Popov's hydrodynamic theory, the bosonic field operator $\Psi (x)$ is written in the hydrodynamic variables, i.e., $\Psi (x) = \sqrt{n_{0} + \pi (x)} e^{i \varphi (x)}$ with $x = ({\bf r}, \tau)$. Here, $\pi (x)$ and $\varphi (x)$ are density and phase fluctuation operators. Green's functions in the hydrodynamic picture and the standard picture are related each other~\cite{Popov1972,Popov1979,Popov1983BOOK,Dupuis2011,Dupuis2011LP}, giving the form 
\begin{align}
G (p) = & - \frac{1}{4 n_{0}} G_{\pi\pi} (p) \begin{pmatrix} 1 & 1 \\ 1 & 1 \end{pmatrix} + i G_{\pi \varphi} (p) \begin{pmatrix} 1 & 0 \\ 0 & - 1 \end{pmatrix}
\nonumber \\ 
& - n_{0} G_{\varphi\varphi} (p) \begin{pmatrix} 1 & -1 \\ -1 & 1 \end{pmatrix} + \delta G (p) \begin{pmatrix} 1 & 1 \\ 1 & 1 \end{pmatrix}, 
\label{correGPopov}
\end{align} 
where 
\begin{align}
\delta G (p) = - \frac{n_{0}}{2} T \sum\limits_{q} G_{\varphi\varphi} (p+q) G_{\varphi\varphi} (q) . 
\label{correPopov}
\end{align} 
Here, $G_{AB}$ with $A, B = \pi, \varphi$ is the non-perturbed correlation function for the hydrodynamic variables, given by~\cite{Popov1972,Popov1979,Popov1983BOOK,Dupuis2011,Dupuis2011LP} 
\begin{align}
\begin{pmatrix} G_{\pi\pi} (p) & G_{\pi \varphi} (p) \\ G_{\varphi \pi} (p) & G_{\varphi \varphi} (p) \end{pmatrix} 
 = 
\frac{1}{\omega_{n}^{2} + c_{0}^{2} {\bf p}^{2}} 
\begin{pmatrix}  n_{0}{\bf p}^{2}/m & - \omega_{n} \\ \omega_{n} & m c_{0}^{2} / n_{0} \end{pmatrix}. 
\label{nnrhorho}
\end{align} 
In (\ref{nnrhorho}), we used the conditions (a) and (b), which leads an approximate equality between the mean density and the condensate density, i.e., $n \simeq n_{0}$~\cite{Dupuis2011}. 
\par 
To obtain (\ref{correGPopov}), the bosonic field operator $\Psi (x)$ is expanded by the fluctuation operators $\pi (x)$ and $\varphi (x)$. Fluctuations are considered up to the second-order. According to (\ref{nnrhorho}), the phase fluctuation is stronger than the density fluctuation. Thus, the phase fluctuation effect alone is taken as the second-order fluctuation. 
\par 
In the hydrodynamic regime $|{\bf p}| \ll \sqrt{2} m c_{0} $, by using (\ref{nnrhorho}), we end with 
\begin{align}
G (p) = - \frac{ m c_{0}^{2}}{\omega_{n}^{2} + c_{0}^{2} {\bf p}^{2}} \begin{pmatrix} 1 & -1 \\ -1 & 1 \end{pmatrix} + \delta G (p) \begin{pmatrix} 1 & 1 \\ 1 & 1 \end{pmatrix}. 
\label{hydroG}
\end{align} 
The first term in (\ref{hydroG}) is equivalent to the first term in (\ref{hydroG1}). The second term in (\ref{hydroG}) becomes also equal to the second term in (\ref{hydroG1}) for small $p$. Indeed, in the hydrodynamic regime, a relation $g_{12} (p) \simeq n_{0} G_{\varphi \varphi} (p)$ holds. As a result, $\delta G (p)$ in (\ref{correPopov}) is reduced into $\Pi_{14} (p)/(2n_{0})$, which reproduces the second term in (\ref{hydroG1}). 
\par 
To summarize, in the hydrodynamic regime, the Green's function in our approach (\ref{eq.intro5}) reproduces the Green's function obtained in the Popov's hydrodynamic approach. In particular, the term (\ref{correPopov}) including the convolution of the phase-phase correlation is reproduced from the second term in (\ref{eq.intro5}), where $\tilde \Sigma$ involves the single bubble self-energy (\ref{eq4}). 
\par 
One of the highlights in the Popov's hydrodynamic approach is to meet the NN identity~\cite{Popov1979}. We substitute the Green's function (\ref{hydroG}) into the Dyson--Beliaev equation. Inversely solving this Dyson--Beliaev equation $\Sigma (p) = G_{0}^{-1} (p) - G^{-1} (p)$, we obtain $\Sigma (0) = {\bf \mu}$. This result meets the Hugenholtz-Pines relation~\cite{Hugenholtz1959} as well as the NN identity~\cite{Nepomnyashchii1975,Nepomnyashchii1978}. This equality $\Sigma (0) = {\bf \mu}$ originates from the fact that the term $\delta G$ has the infrared divergence. Our Green's function approach employs the same procedure to obtain (\ref{Eq214}). 
\par  
\section{Nepomnyashchii--Nepomnyashchii identity}\label{AppendixC}
\par 
\subsection{Derivation of Eq.~(\ref{2-3relation})}\label{AppendixC1}
\par 
To derive the equality (\ref{2-3relation}), it is convenient to refer to an exact many-line vertex $M ( r_{\rm out}, r_{\rm in}, r_{U})$, given by Nepomnyashchii and Nepomnyashchii~\cite{Nepomnyashchii1978}. Here, $r_{\rm in}$ and $r_{\rm out}$ are numbers of incoming and outgoing external particle lines, respectively. $r_{U}$ is the number of an external potential line $U$. In this vertex $M$, momentum and frequency are taken to be zeros with respect to the external particle line and the external interaction potential line. 
\par 
The exact many-line vertex, which is constructed from diagrams irreducible in the particle lines, reads as~\cite{Nepomnyashchii1978} 
\begin{align} 
M ( r_{\rm out}, r_{\rm in}, r_{U} ) 
= & n_{0}^{(r_{\rm out} - r_{\rm in}) /2 } \left ( - \frac{\partial }{\partial \mu} \right )_{n_{0}}^{r_{U}} \left (  \frac{\partial }{\partial n_{0}} \right )_{\mu}^{r_{\rm out}}  
n_{0}^{r_{\rm in}} \left (  \frac{\partial }{\partial n_{0}} \right )_{\mu}^{r_{\rm in}} E' (T, \mu, n_{0}), 
\label{EMLV}
\end{align} 
where $E'$ is the thermodynamic potential given by $E' = - T \ln {\rm Tr}[\exp ({-\beta H'})]$. For the Hamiltonian $H'$, we subtract the contribution $- \mu n_{0}$ from the original Hamiltonian (\ref{eq1}), where the Bogoliubov prescription is applied. Indeed, we are considering diagrams irreducible in the particle lines. An operator $\partial / \partial \mu$ creates a vertex point connecting to an external potential line $U$. An operator $ \sqrt{n_{0}} \partial / \partial n_{0}$ creates a vertex point connecting to an external particle line by eliminating one condensate line. 
\par 
Using (\ref{EMLV}), we obtain matrix forms of the two-point vertex (the self-energy) $\Sigma$ and the three-point vertex $P^{\dag}$ with respect to the external particle line, which are respectively given by~\cite{Nepomnyashchii1978} 
\begin{eqnarray}
\Sigma (0) =  \frac{\partial E'}{\partial n_{0}} \begin{pmatrix} 1 & 0 \\ 0 & 1 \end{pmatrix} + n_{0} \frac{\partial^{2} E'}{\partial n_{0}^{2}} \begin{pmatrix} 1 & 1 \\ 1 & 1 \end{pmatrix} , 
\label{NNbfSigma}
\end{eqnarray}
\begin{eqnarray}
P^{\dag} (0;0) = 2 \sqrt{n_{0}} \frac{\partial^{2} E'}{\partial n_{0}^{2}} \eta^{\dag} + n_{0}^{3/2} \frac{\partial^{3} E'}{\partial n_{0}^{3}} \eta_{\rm a}^{\dag} . 
\label{langlePNN}
\end{eqnarray} 
We thus obtain the relation (\ref{2-3relation}). 
\par 
We note that the Hugenholtz-Pines relation $\mu = \Sigma_{11} (0) - \Sigma_{12} (0)$ is also obtained from (\ref{NNbfSigma}), when we apply $\mu = \partial E' / \partial n_{0}$~\cite{Nepomnyashchii1978}. 
\par 
\subsection{Original derivation of Nepomnyashchii--Nepomnyashchii identity}\label{AppendixC2}
\par  
\begin{figure}
\begin{center}
\includegraphics[width=8.5cm]{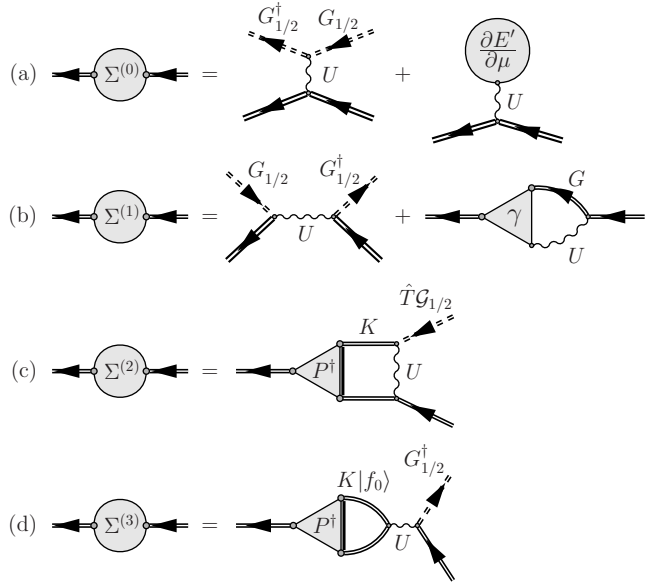}
\end{center}
\caption{Self-energy $\Sigma = \Sigma^{(0)} + \Sigma^{(1)} + \Sigma^{(2)} + \Sigma^{(3)}$ studied by Nepomnyashchii and Nepomnyashchii~\cite{Nepomnyashchii1978}, where all diagrammatic contributions are included. Diagrams in (a), (b), (c) and (d) correspond to $\Sigma^{(0)}$, $\Sigma^{(1)}$, $\Sigma^{(2)}$, and $\Sigma^{(3)}$, respectively.}
\label{fig9.fig}
\end{figure} 
\par   
Nepomnyashchii and Nepomnyashchii considered the full self-energy contribution by using vertex functions~\cite{Nepomnyashchii1978} (diagrammatically described in Fig.~\ref{fig9.fig}), giving the form 
\begin{align}
\Sigma (p) = & \Sigma^{(0)} (p) + \Sigma^{(1)} (p) + \Sigma^{(2)} (p) + \Sigma^{(3)} (p), 
\end{align}
where 
\begin{align}
\Sigma^{(0)} (p) =  & - U \frac{1}{2} G_{1/2}^{\dag} G_{1/2} - U \frac{\partial E'}{\partial \mu} = U n,  
\label{SigmaNN0}
\\
\Sigma^{(1)} (p) =  & - G_{1/2} U G_{1/2}^{\dag} - T \sum\limits_{q} U \gamma (q;p) G (q) , 
\label{SigmaNN1}
\\
\Sigma^{(2)} (p) = & - T \sum\limits_{q} P^{\dag} (q;p) K_{} (q;p) U \frac{ 1 }{ \sqrt{-1}  } \hat T {\mathcal G}_{1/2} , 
\label{SigmaNN2}
\\
\Sigma^{(3)} (p) = & - \frac{T}{2} \sum\limits_{q} P^{\dag} (q;p) K_{} (q;p) | f_{0}\rangle U \frac{1}{\sqrt{-1}} G_{1/2}^{\dag}. 
\label{SigmaNN3}
\end{align} 
Here, $K_{} (q;p)$ is a bare part of the $(4\times 4)$-matrix two-particle Green's function, given by 
\begin{align}
K (p+q) = G (p+q) \otimes G(-q). 
\label{KfullG}
\end{align} 
The Green's function $G_{}$ is the full Green's function, where all the diagrammatic contributions are included to the self-energy. In the small-$p$ regime, the leading term of $G_{}$ is reduced to~\cite{Gavoret1964,Nepomnyashchii1978} 
\begin{align}
G (p)= & - \frac{n_{0} m c^{2}}{n} \frac{1}{\omega_{n}^{2} + c^{2} |{\bf p}|^{2}} \begin{pmatrix} 1 & -1 \\ - 1 & 1 \end{pmatrix}. 
\label{fullG}
\end{align} 
In (\ref{SigmaNN1}), $\gamma (p;q)$ is the $(2\times 2)$-matrix three point vertex that has an external potential line and two external particle lines. In (\ref{SigmaNN0}), $\partial E' / \partial \mu$ corresponds to the one point vertex connecting to an external potential line. We used the fact that this vertex is equivalent to the non-condensate density, i.e., $n' = - \partial E' / \partial \mu$. 
\par 
The contributions (\ref{SigmaNN0}) and (\ref{SigmaNN1}) converge. On the other hand, for (\ref{SigmaNN2}) and (\ref{SigmaNN3}), the bare part of the two-particle Green's function $K$ provides the infrared divergences. Since the infrared divergences are strongly related to each other, these are simply extracted by using $K^{\rm IR} (q;p) = G_{12} (p+q) G_{12} (q) \hat C$, where we used the symmetry relation $G_{12} (p) = G_{12} (-p)$. Indeed, according to (\ref{fullG}), we have the infrared-divergent relation $\lim_{p\rightarrow 0} G_{11,22} (p) = - \lim_{p\rightarrow 0} G_{12,21} (p)$~\cite{Gavoret1964}. 
\par 
The infrared divergences in (\ref{SigmaNN2}) are exactly canceled out, because we have a relation $\hat C \hat T \hat \eta_{\rm g} = 0$, as discussed in the case of the first term of (\ref{SigmaIRvc}). On the other hand, for (\ref{SigmaNN3}), the infrared divergence remains as discussed in the second term in (\ref{SigmaIRvc}).

Using the identity (\ref{2-3relation}) as well as the second mean value theorem for integrals~\cite{Iyanaga1977}, we can reduce the self-energy $\Sigma$ to 
\begin{align}
\Sigma (0) = & 2 U \Pi_{\rm IR}^{\Lambda} \Sigma_{12} (0) \begin{pmatrix} 1 & 1 \\ 1 & 1 \end{pmatrix} + \delta \Sigma
\label{Eq148}
\end{align} 
Here, $\Pi_{\rm IR}^{\Lambda} \equiv - T \sum_{{\bf q}=0}^{\Lambda} [ G_{12} (i \omega_{n} = 0, {\bf q}) ]^{2}$ provides the infrared divergence. The $(2\times 2)$-matrix $\delta \Sigma$ is a remaining converging part of $\Sigma$. Solving (\ref{Eq148}) for $\Sigma_{12} (0)$, we end with $\Sigma_{12} (0) = 0$, where we used $\Pi_{\rm IR}^{\Lambda} = \infty$. This is the original derivation of the NN identity. 
\par 
To summarize, the NN identity was originally derived from (a) the infrared divergence of the Green's function due to the spontaneously broken continuous symmetry, as well as (b) the Ward-Takahashi identity with respect to two and three-point vertices. The diagrammatic contribution essential to the NN identity is the bubble structure diagram with the vertex correction (\ref{SigmaNN3}), which is the same diagram as the second term of (\ref{SigmaIRvc}) in our formalism. 
\par 
\subsection{Gauge invariance}\label{AppendixC3}
\par 
To derive the NN identity, Nepomnyashchii and Nepomnyashchii used the fact that the Green's function exhibits the infrared divergence, which originates from the phase fluctuation thanks to the spontaneously broken continuous gauge symmetry. If we apply the idea that physical quantities are gauge invariant, we may more simply understand the NN identity as well as the finiteness of the chemical potential $\mu$, the diagonal self-energy $\Sigma_{11}$, and the macroscopic sound speed $c$. 
\par 
We apply the exact many-line vertex (\ref{EMLV}) at $p=0$. Note that if the system is not at the critical temperature, the exact many-line vertex $M$ does not diverge. Indeed, it is given by thermodynamic derivative of the condensate density and the chemical potential. This vertex $M$ is thus finite not at $T_{\rm c}$. We also note that the thermodynamic potential $E'$ is gauge invariant. In the representation (\ref{EMLV}), we assumed that the BEC order parameter is a real number. We now consider that the BEC order parameter is a complex number, given by $\sqrt{n_{0}} e^{i \varphi_{0}}$. In this case, the exact many-line vertex (\ref{EMLV}) has the factor given by $\exp{[ i \varphi_{0} (r_{\rm out} - r_{\rm in}) ]}$. 
\par 
The off-diagonal self-energy $\Sigma_{12}$ at $p=0$ is now given by 
\begin{align}
\Sigma_{12} (0) = e^{2 i \varphi_{0} }n_{0} \frac{\partial^{2} E'}{\partial n_{0}^{2}}. 
\label{NNSigma12}
\end{align} 
The phase $\varphi_{0}$ is chosen to be arbitrary in the ordered phase with the spontaneously broken continuous symmetry. We thus take the average with respect to $\varphi_{0}$ over the range $\varphi_{0} \in [0, 2 \pi)$. The averaged off-diagonal self-energy $\langle \Sigma_{12} (0) \rangle$ is now given by $\langle \Sigma_{12} (0) \rangle = 0$, which leads to the NN identity. 
\par 
On the other hand, the diagonal self-energy is given by 
\begin{align}
\Sigma_{11} (0) = & \frac{\partial E'}{\partial n_{0}} + n_{0} \frac{\partial^{2} E'}{\partial n_{0}^{2}}. 
\end{align} 
The chemical potential is given by $\mu = \partial E' / \partial n_{0}$~\cite{Nepomnyashchii1978}. The density-density correlation function $\chi_{}$, which has two external potential lines, reads as 
\begin{align}
\chi_{} (0) = & \frac{\partial^{2} E'}{\partial \mu^{2}} = - \frac{\partial n'}{\partial \mu}. 
\end{align} 
As shown in Ref.~\cite{Nepomnyashchii1978}, we have $\chi (0) = - n /(mc^{2})$. These quantities do not involve a factor given by $\exp{(i \varphi_{0})}$, and their values are unchanged even if we take the average with respect to the phase $\varphi_{0}$. 
\par 
To summarize, averaged quantities of gauge invariant operators are not affected by uncertainty of the phase $\varphi_{0}$. The gauge invariance protects their finiteness against the phase fluctuations. On the other hand, the gauge-dependent quantities are affected by the phase fluctuations, and their values vanish at $p=0$. It may lead to the NN identity $\Sigma_{12} (0) = 0$~\cite{Nepomnyashchii1975,Nepomnyashchii1978}. 
\par 
 

\begin{thebibliography}{99}
\bibitem{note2} In these theories, $\Gamma_{11}^{\rm R}$ in (\ref{Eq243}) and (\ref{Eq244}) is replaced by $\Gamma_{11}$ in (\ref{RPABetheSalpeter1}), and $U_{\rm eff}^{\rm R}$ in (\ref{eq27}) and (\ref{eq28}) is replaced by $U_{\rm eff}$.
\bibitem{note3} We note that $\Pi_{14}(p\to 0)$ is independent of $n_{0}$ when $T>0$, as seen in Eq.~(\ref{AppBlast}).
\end{thebibliography}
\end{document}